\documentclass[11pt]{article}

\usepackage[printonlyused]{acronym}
\usepackage{authblk}
\usepackage{amsmath}
\usepackage{booktabs}
\usepackage{caption}
\usepackage{dirtytalk}
\usepackage[shortlabels]{enumitem}
\usepackage{float}
\usepackage[hang,flushmargin]{footmisc}
\usepackage[a4paper, margin=2cm]{geometry}
\usepackage{graphicx}
\usepackage{hyperref}
\usepackage[utf8]{inputenc}
\usepackage{listings}
\usepackage{multirow}
\usepackage{setspace}
\usepackage{subcaption}
\usepackage{wasysym}
\usepackage{xltabular}

\hypersetup{
    colorlinks,
    linkcolor={black},
    citecolor={black},
    urlcolor={black}
}

\graphicspath{{./figures/}}

\newcommand{\seeref}[1]{%
    (see \autoref{#1})%
}

\title{Evaluation of a Data Annotation Platform for Large, Time-Series Datasets in Intensive Care: Mixed Methods Study}

\author[1,2]{Marceli Wac}
\author[1]{Raul Santos-Rodriguez}
\author[1,2]{Chris McWilliams}
\author[2]{Christopher Bourdeaux}

\affil[1]{Faculty of Engineering, University of Bristol}
\affil[2]{University Hospitals Bristol and Weston NHS Foundation Trust}

\date{September 2023}

\begin{document}
\maketitle

\section{Introduction}

\acp{ICU} are complex and data-rich environments where critically ill patients are admitted due to the seriousness of their condition \cite{bennett_organisation_1999}, frequently necessitating multiple organ support, continuous observations and sophisticated treatment such as mechanical ventilation.
\acp{ICU} are characterised by a large number of medical devices collecting data surrounding patient's treatment, medication intake and vital signs.
This routinely collected data is collected by and aggregated in a clinical information system which allows clinicians to access it to inform the provision of care.
Including vital signs, medication, laboratory test results and treatment-specific measurements - each patient can generate over 1000 different types of data points over the course of their stay in the \ac{ICU} \cite{johnson_mimic-iv_2022}, resulting in a large volume of data which is difficult to manage and utilise effectively.

The application of \ac{ML} and computational techniques within healthcare is a continuously growing area of interest within both research and commercial settings, and allows for more effective utilisation of the data to target specific challenges and problems within healthcare \cite{noorbakhsh-sabet_artificial_2019,davenport_potential_2019}.
In a typical \ac{ML} workflow within healthcare, a subset of the collected data is used together with an \ac{ML} algorithm to compute, or train a \ac{ML} model capable of making predictions or deriving insights from the data.
Such model can then be deployed within the clinical setting where it continuously ingests the patient data and integrates with the clinical information systems to deliver insights to the clinical team and help them stratify their patients, inform their decisions or predict adverse events before they become apparent to the staff.

While the data gathered in the \acp{ICU} can often be used directly, for example, to present the correlation between different vital signs and the patient prognosis, complex tasks that rely heavily on clinical experience and expertise may require human involvement to provide further guidance \cite{hunter_role_2022}.
This guidance, most frequently referred to as labels or annotations \cite{heo_artificial_2021}, can deliver additional information or context to the existing data.
An example of such a label could be a \say{yes} or \say{no} indicator of whether a patient is ready for discharge or the type and class of tumour present on an x-ray image.
With this further knowledge, \ac{ML} models can take advantage of the human experience to tackle complex problems and deliver solutions that could not be inferred from the raw data alone \cite{azizi_big_2021}.
While the large volumes of data available in intensive care offer significant opportunities, working with this data also presents a unique set of challenges, as the effort required to annotate the dataset increases proportionally to its size \cite{martinez-martin_ethical_2021}.
Furthermore, the complex nature of the healthcare data and the fact that applying a single label can require clinicians to look at multiple parameters, patient history and lab results make the annotation task highly labour-intensive.
With the clinician's time being a remarkably valuable resource, ensuring the effectiveness of the data annotation workflow is paramount.
The lack of an annotation system tailored to the unique nature of the healthcare data (e.g. accessing a subset of relevant variables from a list of potentially hundreds of parameters \cite{burki_artificial_2021}) further complicates this problem.


\subsection{Semi-supervised and Weak Supervision Methods}

Annotation of clinical data is a particularly challenging problem which relies on the domain expertise and frequently large time commitment from its annotators.
This makes traditional approaches, such as \say{crowdsourcing}, impractical due to the large volume of data that needs to be annotated and relatively narrow pool of annotators who have the time and expertise needed to annotate it.
Furthermore, because these approaches assume a regime in which a large number of annotators label small subsets of the dataset, their results frequently rely on the robustness of individual labels and high degree of agreement between the annotators.
The results produced by crowdsourcing may therefore generate data insufficient for training \ac{ML} models using purely supervised approaches (e.g. those relying solely on the labelled data), particularly if the number and diversity of annotations do not adequately cover the variance of data present within the dataset.

Problems addressable by supervised \ac{ML} methods can however utilise additional techniques when the underlying dataset contains only a partially annotated data \cite{poyiadzi_weak_2022}.
These approaches could be categorised as either semi-supervised learning, in which unlabelled data is used to improve the generalisability of the model trained on the limited number of labels, or weak supervision methods, where less accurate or noisy data sources are used to generate labels for the unlabelled data.
Both semi-supervised and weak supervision approaches are particularly well suited to the problems where annotation of the entire dataset is prohibitively difficult or expensive, but obtaining large volumes of unlabelled data is relatively easy \cite{ratner_training_2019}.

One prominent example of weak supervision techniques which could be facilitated using our tool is known as \say{data programming} \cite{ratner_data_2016}.
In data programming, instead of applying labels directly to the data, annotators describe the process through which the labels could be applied, by defining heuristic rules known as labelling functions \cite{ratner_data_2016}.
Depending on the task, these functions can utilise several approaches that rely on existing knowledge bases and libraries, specific characteristics and patterns in the data or the combination of both.
Because each of the labelling functions can prioritise different heuristic and aspect of data, annotations created by individual functions are typically very noisy and biased towards their specific approach or a subset of targeted cases.
Additionally, combining multiple labelling functions is a complex task that needs to address conflicts or overlaps of created labels and the dependencies between the different functions.
This suggests that sets of labelling functions need to undergo additional processing before they can be used to annotate the data.
Several approaches could be assumed for this task, including simple techniques such as majority voting where the annotation is applied only when majority of the labelling functions agree on the label.
These techniques, however, operate under an assumption that all annotators share the same degree of expertise, which is frequently not the case within the clinical setting.
More complex methods of aggregating labels from multiple sources such as multi-task weak supervision (which estimates the accuracy of each labelling function, aggregates their output and produces a model that can be used to annotate the data \cite{ratner_training_2019, snorkel_team_labelmodel_2020}) or the probabilistic approach proposed by Reykar et al. (which \say{jointly learns the classifier/regressor, the annotator accuracy, and the actual true label} \cite{raykar_learning_2010}) have been shown to outperform majority voting and provide more accurate results \cite{ratner_training_2019, raykar_learning_2010}.

\subsection{Research Scope}

In our previous study, we conducted an experimental activity with the clinical staff from the \acp{ICU}, which focused on capturing the approach to the data annotation task \cite{wac_capturing_2023}.
Through the analysis of the observations made during this activity, we established 11 key requirements for a data annotation platform tailored to the large, time-series datasets in the clinical setting \cite{wac_capturing_2023}.
We then analysed existing data annotation tools and evaluated their features for use in the clinical context, and proposed a novel solution to the data annotation in a form of a bespoke tool designed to meet these requirements \cite{wac_cats_2023}.

In this study, we present a novel tool for data annotation within the clinical setting, focusing on two distinct approaches it facilitates.
The first approach allows for direct annotation of individual admissions, while the second allows the annotators to define rulesets that can be used to create labels across the entire dataset \seeref{sec:wls_3_data_collection}.
Both approaches are capable of collecting the annotations from multiple users, and in the case of annotation of individual admissions, our tool also facilitates dynamic assignment of subsets of data to the annotators.
This makes our platform suitable for capturing annotations within multi-user frameworks such as crowdsourcing, but it also enables the use of weak supervision techniques such as data programming by treating collected rulesets as labelling functions.
Furthermore, our tool allows for simultaneous and flexible annotation using either method, making it an adequate solution for the deployment of semi-supervised techniques that supplement partially annotated datasets with additional information from the rulesets.

Designing machine learning pipelines is however a complex process that encompasses various areas beyond the data annotation, such as data pre-processing, model selection, feature engineering and hyper-parameter tuning.
Similarly, the process of data annotation is itself concerned with many challenges such as aggregating the annotations from multiple sources, establishing their biases and accuracies and consolidating multiple labelling functions into a reliable annotation system.
Because each of these areas requires a careful consideration and expertise, they all merit their own dedicated research, which falls outside of the scope of this study.
Instead, in our work we focus solely on the data collection and usability aspects of our tool and acknowledge that further processing of the annotations collected as part of this study may be required to establish a robust annotation model and highlight the limited immediate usability of the annotations generated using the individual rulesets.

\section{Methods}

To evaluate our tool, we set out to capture several metrics which could inform both the usability of our tool and its performance in the data collection as part of the annotation workflow.
We assumed a sequential, mixed-method study design involving practical data annotation with the clinical staff followed by an assessment of its usability.
The study was split into two consecutive stages and involved annotation of weaning from mechanical ventilation.
The \ac{S1} focused on the annotation of individual admissions, in which participants created labels directly on top of the admission data, while the \ac{S2} was centred around the semi-automated approach to annotation and involved participants creating rulesets that would be used to annotate the entire dataset simultaneously.
Both stages were piloted with an \ac{ICU} Consultant who had extensive experience in critical care medicine, as well as researchers from a \ac{ML} and data scientists background prior to the recruitment of participants.
Piloting the activity allowed us to refine the user interface of the software and expand the creation form with additional fields which enabled us to capture the annotator's confidence in the created label and provide a set of parameters that suggested the label creation in the first place.
Both stages were facilitated using the same instance of our custom tool developed for this study \cite{wac_cats_2023}.

\subsection{Participant Recruitment}

The recruitment of participants for both \ac{S1} and \ac{S2} followed a mixture of convenience and snowball sampling and took place at the \ac{UHBW} in the \ac{UK}.
Participants were invited to each stage of the study separately via e-mail sent to the staff working in the \ac{ICU} who had prior experience with mechanical ventilation treatment.
The e-mails contained the participant information sheet which detailed the purpose of the study, instructions for the enrolment using our tool as well as an illustrated guide of the annotation process facilitated using our tool. 
The enrolment process was facilitated entirely through our tool and involved participants visiting a provided link, which presented them with an account creation form and a link to an electronic signature-enabled consent form.
Upon completion of the sign-up process, participants had to verify both their electronic signature and an account created on our platform using links sent to their e-mail address.
Crucially, as \ac{S1} and \ac{S2} were recruited for separately, participants who took part in \ac{S1} were not required to, or excluded from taking part in \ac{S2}.

\begin{figure}
  \includegraphics[width=\textwidth]{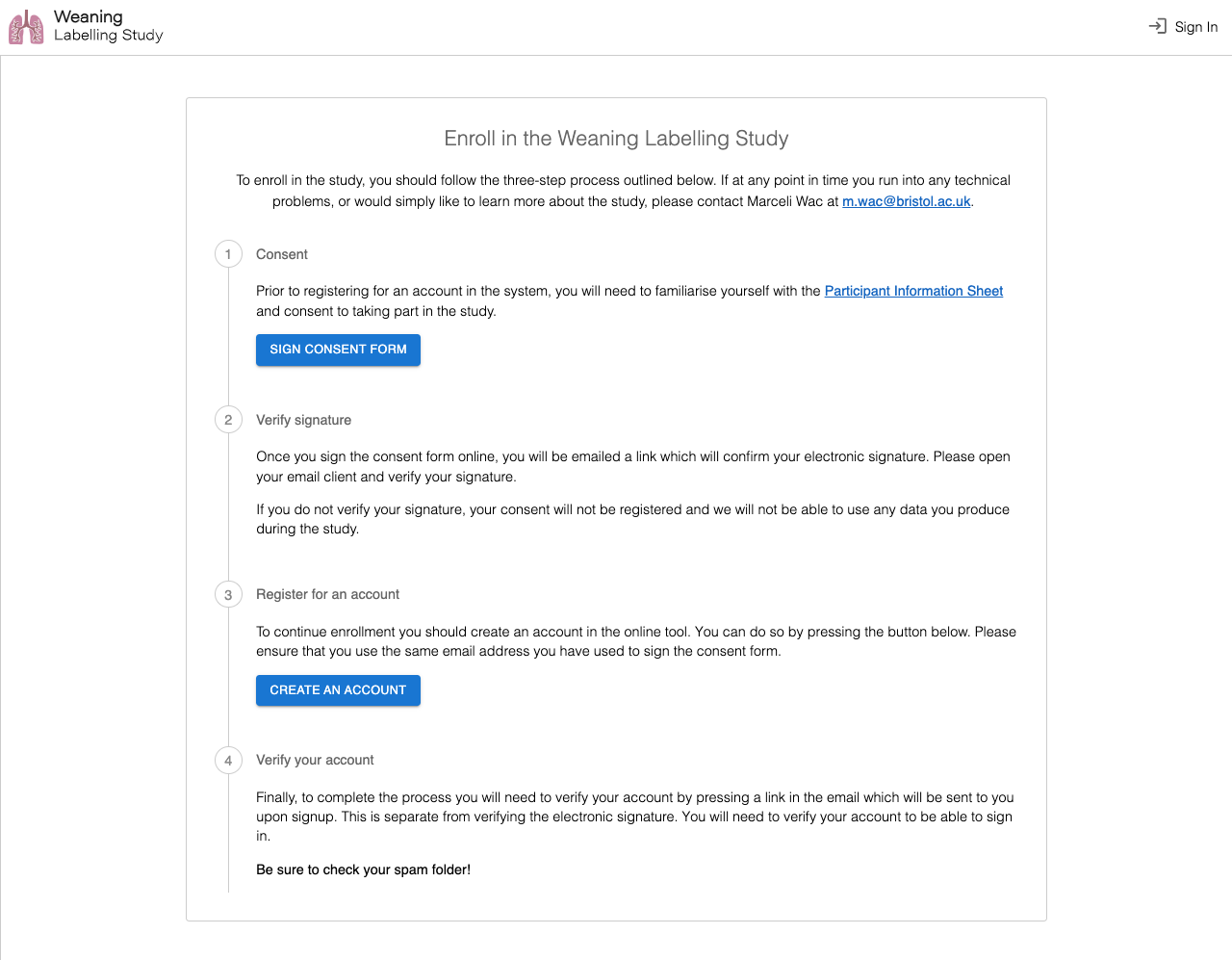}
  \caption{Enrolment screen embedded within our tool to guide participants on how to take part in our study.}
  \label{fig:enrollment}
\end{figure}

We encountered significant challenges in the recruitment and engagement of participants throughout the course of this study.
This necessitated the development of interventions which would increase the number of participants and promote involvement in the research activities.
Section \ref{sec:weaning_labelling_evaluation_recruitment} outlines the problems encountered during the participant recruitment and engagement throughout the study, as well as the mitigation strategies implemented to assimilate them.
Our recruitment process yielded a sample size of 28 participants, of whom 22 consented to take part in \ac{S1} and 13 consented to take part in \ac{S2}.
Of those, 12 participants engaged with \ac{S1} by creating at least 1 annotation and 9 engaged with \ac{S2} by creating at least 1 ruleset; 1 of the participants engaged in annotation in both \ac{S1} and \ac{S2} and 8 never engaged with annotation despite consenting to take part in the study \seeref{table:wls_3_participants}.

The research site (\ac{UHBW}) had a pre-established research partnership with University of Bristol and some of the participants might have known the researchers involved through their professional relationship and the research previously undertaken at this site.

\begin{table}
\setstretch{1.25}
\caption{Participants and their job roles aggregated across the two stages of the study. Columns \textit{enrolled} and \textit{annotated} denote whether participant consented to take part in the study and created at least one annotation in the given stage of the study respectively. \label{table:wls_3_participants}}
\begin{tabularx}{\linewidth}{
    |c|X|c|c|c|c|
}
    
    \cline{3-6}
    \multicolumn{2}{l|}{} & \multicolumn{2}{c|}{\textbf{Enrolled}} & \multicolumn{2}{c|}{\textbf{Annotated}} \\
    \hline
    \textbf{ID} & \textbf{Job role} & \textbf{Stage 1} & \textbf{Stage 2} & \textbf{Stage 1} & \textbf{Stage 2} \\
    \hline
    p01 & \ac{ICU} Consultant & \XBox & \XBox & \XBox & \XBox \\
    \hline
    p02 & \ac{ICU} Consultant & \Square & \XBox & \Square & \XBox \\
    \hline
    p03 & \ac{ICU} Registrar & \Square & \XBox & \Square & \XBox \\
    \hline
    p04 & \ac{ICU} Consultant & \XBox & \XBox & \Square & \XBox \\
    \hline
    p05 & \ac{ICU} Consultant & \XBox & \Square & \XBox & \Square \\
    \hline
    p06 & \ac{ICU} Registrar & \XBox & \Square & \XBox & \Square \\
    \hline
    p07 & \ac{ICU} Registrar & \XBox & \Square & \XBox & \Square \\
    \hline
    p08 & \ac{ICU} Consultant & \XBox & \Square & \XBox & \Square \\
    \hline
    p09 & Paediatric ICU Consultant & \Square & \XBox & \Square & \XBox \\
    \hline
    p10 & \ac{ICU} Registrar & \Square & \XBox & \Square & \XBox \\
    \hline
    p11 & \ac{ICU} Fellow & \XBox & \XBox & \Square & \Square \\
    \hline
    p12 & \ac{ICU} Fellow & \XBox & \Square & \XBox & \Square \\
    \hline
    p13 & \ac{ICU} Fellow & \XBox & \Square & \XBox & \Square \\
    \hline
    p14 & \ac{ICU} Consultant & \XBox & \Square & \XBox & \Square \\
    \hline
    p15 & \ac{ICU} Registrar & \XBox & \XBox & \Square & \Square \\
    \hline
    p16 & \ac{ICU} Consultant & \XBox & \Square & \XBox & \Square \\
    \hline
    p17 & Advanced Critical Care Practitioner & \XBox & \XBox & \Square & \Square \\
    \hline
    p18 & \ac{ICU} Physiotherapist & \XBox & \XBox & \Square & \Square \\
    \hline
    p19 & \ac{ICU} Consultant & \XBox & \Square & \XBox & \Square \\
    \hline
    p20 & \ac{ICU} Consultant & \XBox & \XBox & \Square & \XBox \\
    \hline
    p21 & \ac{ICU} Registrar & \XBox & \Square & \XBox & \Square \\
    \hline
    p22 & Paediatric ICU Consultant & \Square & \XBox & \Square & \XBox \\
    \hline
    p23 & \ac{ICU} Physiotherapist & \XBox & \Square & \XBox & \Square \\
    \hline
    p24 & Paediatric ICU Consultant & \Square & \XBox & \Square & \XBox \\
    \hline
    p25 & \ac{ICU} Registrar & \XBox & \Square & \Square & \Square \\
    \hline
    p26 & \ac{ICU} Consultant & \XBox & \Square & \Square & \Square \\
    \hline
    p27 & \ac{ICU} Fellow & \XBox & \Square & \Square & \Square \\
    \hline
    p28 & \ac{ICU} Registrar & \XBox & \Square & \Square & \Square \\
    \hline
    \hline
    \multicolumn{2}{|l|}{\textbf{Total count}} & 22 & 13 & 12 & 9 \\
    \hline
\end{tabularx}
\end{table}

\subsection{Dataset for Annotation}
To provide data adequate for the annotation process, \ac{MIMIC} \cite{johnson_mimic-iv_2022} was used as the underlying dataset.
The data was processed to produce a format which could be used in annotation by aggregating time-series parameters relevant to the mechanical ventilation on an hourly basis for each of the eligible admissions.
The included parameters were selected by two independent clinicians working in the \ac{ICU} and consisted of 50 parameters, including vital signs, haemodynamics, and mechanical ventilation-specific parameters \seeref{appendix:parameters}.
The eligibility criteria used to select the admissions required that subjects were at least 18 years old at the time of admission, had undergone an invasive mechanical ventilation treatment that lasted for a minimum of 24 hours during their admission, their stay in the \ac{ICU} lasted for a minimum of 4 days, and that did not end with death, including up to 48 hours following discharge.
The data was extracted using an \ac{SQL} script ran on the PostgreSQL installation of \ac{MIMIC} with the concept tables computed \cite{johnson_mimic_2018}.

\subsection{Data Collection}
\label{sec:wls_3_data_collection}

\subsubsection{Stage 1 - Annotation of Individual Admissions}

The first stage focused on the annotation of individual admissions and used a configuration that assigned 40 admissions to each of the participants.
Of those, 20 were shared among all participants, and 20 were unique to each of the participants.
In order to ensure that the annotated data could be used to compare the annotations created by different participants while maximising the total number of annotated admissions, the assigned admissions were ordered in an alternating fashion (e.g. every other admission annotated by a participant belonged to a \say{shared} set, while the rest were unique to that participant).
\ac{S1} annotation activity was designed to run over the course of 28 days in an entirely asynchronous way, allowing each participant to undertake the task at the time that suited their schedule and over any number of sessions.
Due to the initially poor engagement with the study during this stage, this approach had to have been revised.
While the original time-frame for the data collection remained the same, we conducted several interventions to further promote the study with prospective participants and improve the engagement with the annotation activity (see Section \autoref{sec:weaning_labelling_evaluation_recruitment}).
This also resulted in a change from our initial approach to conducting the annotation activity exclusively asynchronously and remotely to a mixture of in-person, synchronous workshops followed by further asynchronous participation.

Participants of \ac{S1} were instructed to annotate as many admissions as they could over the course of this stage.
The annotation process allowed participants to view the list of admissions assigned to them, view any admission in the list independently and navigate to their next unannotated admission.
Upon annotating an admission, our tool automatically directed the participants to their next assigned admission to preserve the continuity of the workflow.

\subsubsection{Stage 2 - Semi-Automated Annotation}

In \ac{S2}, participants were asked to create a ruleset that best describes the weaning from mechanical ventilation when applied to the entire dataset.
Due to the more technical nature of the task, participants were introduced to the tool during a remote video call which lasted approximately 75 minutes.
Over the course of the call, the facilitator demonstrated the functionality of the software by sharing his screen, creating a ruleset and evaluating it using the calculated statistics which lasted approximately 45 minutes.
The participants were then asked to perform the task on their own and given a chance to ask questions and seek any further clarification on how to use the software over the remaining 30 minutes.
Following this onboarding process, participants were given access to the software and allowed to use it independently over the course of 7 consecutive days at the time that suited their schedules.
Over that time, participants were asked to use the tool to create a ruleset that captures the weaning from mechanical ventilation most accurately to the best of their ability.
Each participant was able to freely experiment with the software and create any number of rulesets.
Following the completion of the 7-day period, participants were asked to provide the identifier of the \say{final} ruleset that marked their best attempt at solving the task, and to fill out a questionnaire capturing their feedback on the usability and feasibility of the data annotation tool \seeref{table:wls_3_questionnaire}.
The choice to utilise questionnaires was made on the basis of the highly focused metrics we aimed to capture and the straightforward nature of collecting data they offered. 
The questionnaire was designed to elicit feedback on several key metrics surrounding the usability of the software and the views on the annotation process as a whole.
These included the overall sentiment to the data annotation process, the ease of use of our tool, specific points of friction and potential improvements to the software, as well as participants' attitudes towards the semi-automated approach to annotation.

\begin{table}
\setstretch{1.25}
\caption{Questions featured in the Stage 2 evaluation questionnaire were used to assess the usability and feasibility of the software and identify potential areas for its improvement.\label{table:wls_3_questionnaire}}
\begin{tabularx}
{\linewidth}{
    |c|X|
}
    \hline
    \textbf{No.} & \textbf{Question} \\
    \hline
    1. & How did you find the data annotation tool overall? \\
    \hline
    2. & Did you find the tool easy or difficult to use? \\
    \hline
    3. & Were there any aspects of activity you have struggled with? \\
    \hline
    4. & Can you think of any improvements that would make this tool or process better? \\
    \hline
    5. & Was there anything you found particularly interesting or challenging in the process of data annotation? \\
    \hline
    6. & Do you think the annotation of data using the ruleset-based approach (compared to manually annotating each admission) is a feasible way of annotating large clinical datasets? \\
    \hline
    7. & Was there anything you discovered or learnt about the weaning from mechanical ventilation by taking part in this activity? \\
    \hline
    8. & Having taken part in this activity, would you change anything about how you approach the problem of weaning from mechanical ventilation in practice? \\
    \hline
\end{tabularx}
\end{table}

\section{Interventions to Promote Study Engagement}
\label{sec:weaning_labelling_evaluation_recruitment}

\begin{figure}[tbp]
  \includegraphics[width=\textwidth]{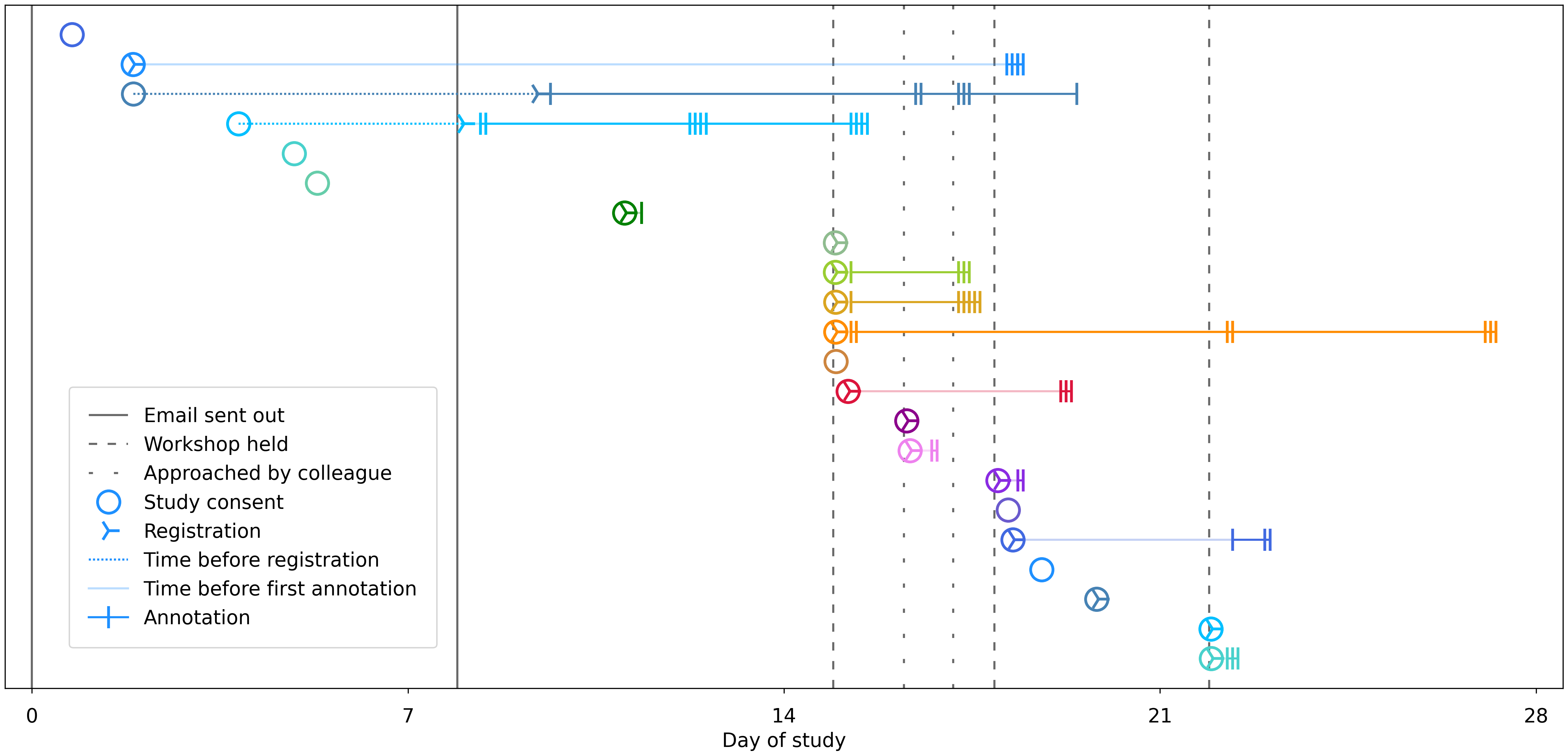}
  \caption{Participant engagement throughout the study has improved following our interventions. Workshops were most effective at recruiting new participants, while the approach by a colleague resulted in the highest increase in the number of created annotations.}
  \label{fig:s1_timeline}
\end{figure}

The initial approach to recruitment for \ac{S1} involved sending out an e-mail with a call for participation.
In addition to the study information, the e-mail explained the enrolment process, in which participants were prompted to visit the attached link to the platform, sign the consent form and create an account, before proceeding to the data annotation task.
This proved ineffective and resulted in 6 consent signatures, 1 account registration and no annotations.
These unsatisfactory results necessitated intervention which would spur the recruitment of additional participants.
Initially, we re-sent the e-mail inviting participants to take part in the study, generating a further 1 consent signature, 3 account registrations and 8 annotations.
While this increased the number of participants for our study, it had a minimal effect on their engagement in the annotation task, indicating the need for further interventions targetting participant engagement specifically.
The limited response to the call for participation suggested that a more personal approach, similar to the in-person workshop held in our previous study \cite{wac_capturing_2023} could prove more feasible.
To further incentivise participation in our study, we decided to host a series of workshops focused on discussing the use of data science within healthcare framed as \say{structured learning events} which are required as part of the trainees' portfolio to progress their careers.
Following the end of the second week of the study, we held three 2-hour workshops which further increased the number of our participants, resulting in an additional 13 consent signatures, 10 registrations and 31 annotations.
While this also improved the engagement of the existing participants demonstrated by the annotations they created following the workshops, participants who enrolled during the workshops did not subsequently engage with the study.
Our final approach to remedy this and encourage more continuous participation focused on utilising the existing community at the hospital.
To that extent, we expanded our research team by onboarding one of the members of clinical staff working at the hospital, to signpost and promote the study among her peers.
This approach proved particularly effective at improving the engagement with the activity, generating further 2 consent signatures, 2 registrations and 17 annotations.
The timeline demonstrating participant engagement and interventions undertaken during \ac{S1} is depicted in the \autoref{fig:s1_timeline}.

To ensure better engagement for \ac{S2}, we applied the insights learnt from our previous study \cite{wac_capturing_2023} and the interventions deployed during \ac{S1}.
Our approach to recruitment included an invitational e-mail to the staff working in the \acp{ICU} in \ac{UHBW}, as well as an in-person approach facilitated by the member of clinical staff incorporated in our team; the participants were further incentivised to enrol in the study with an offer of possibility to co-author the paper upon study completion.
We restructured the activity from an entirely asynchronous and remote process to one that began with an onboarding video call and asynchronous annotation over the following 7 days.
Over the course of three weeks, 12 participants consented to take part in the study, of whom 9 were on-boarded onto the platform and completed the annotation task.
One of the participants suggested an additional person who in their description fitted the candidate profile for the study and was subsequently recruited.

\subsection{Impact of Interventions}

\begin{figure}[tbp]
    \begin{subfigure}[t]{0.48\textwidth}
        \includegraphics[width=\textwidth]{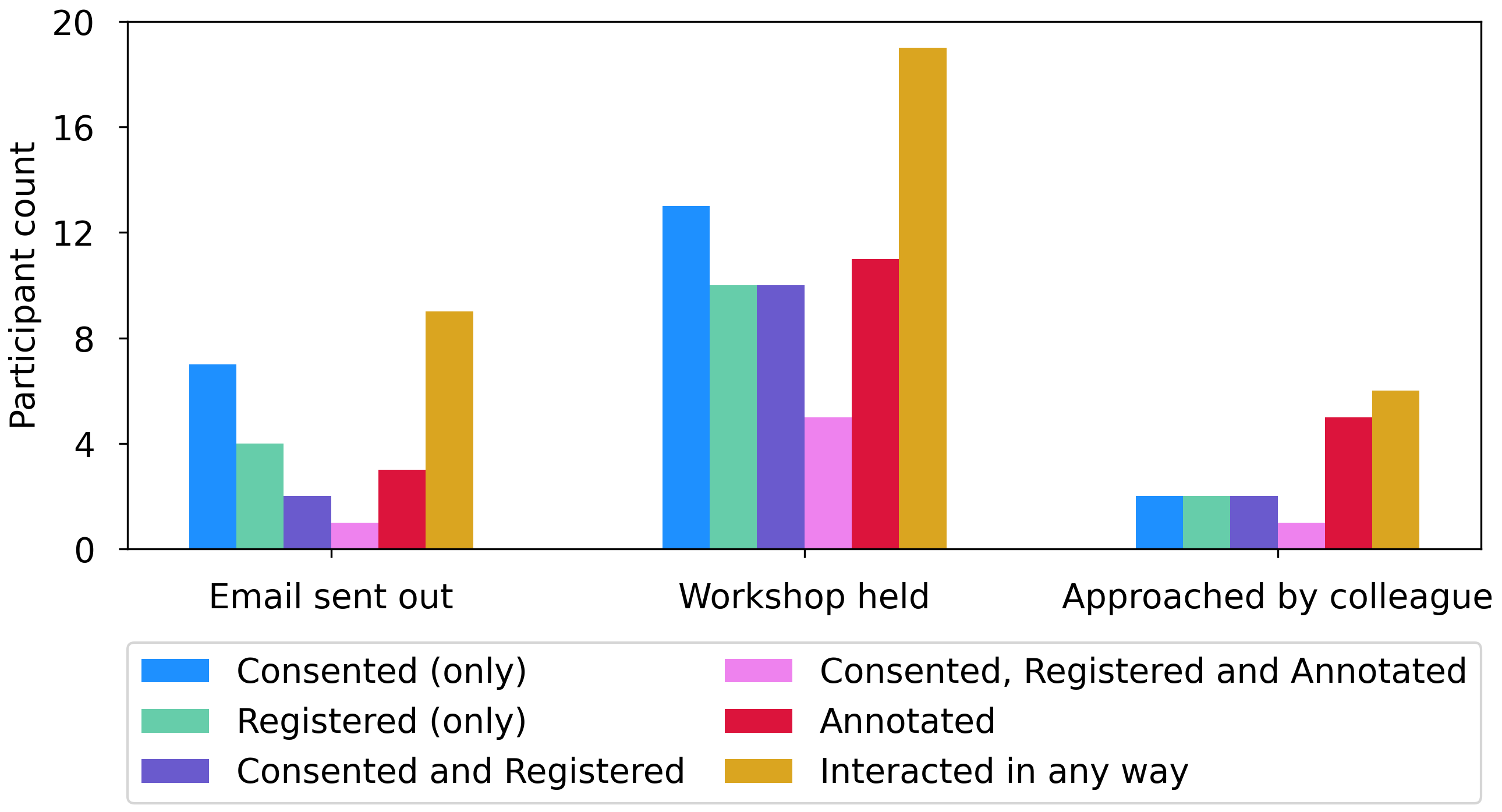}
        \caption{Participants' engagement differed following each intervention. Workshops generated the highest number of consent signatures, registrations and annotating participants; while the approach by the colleague was primarily effective at engaging participants with the annotation activity.}
        \label{fig:s1_interventions}
    \end{subfigure}
    \hfill
    \begin{subfigure}[t]{0.48\textwidth}
        \includegraphics[width=\textwidth]{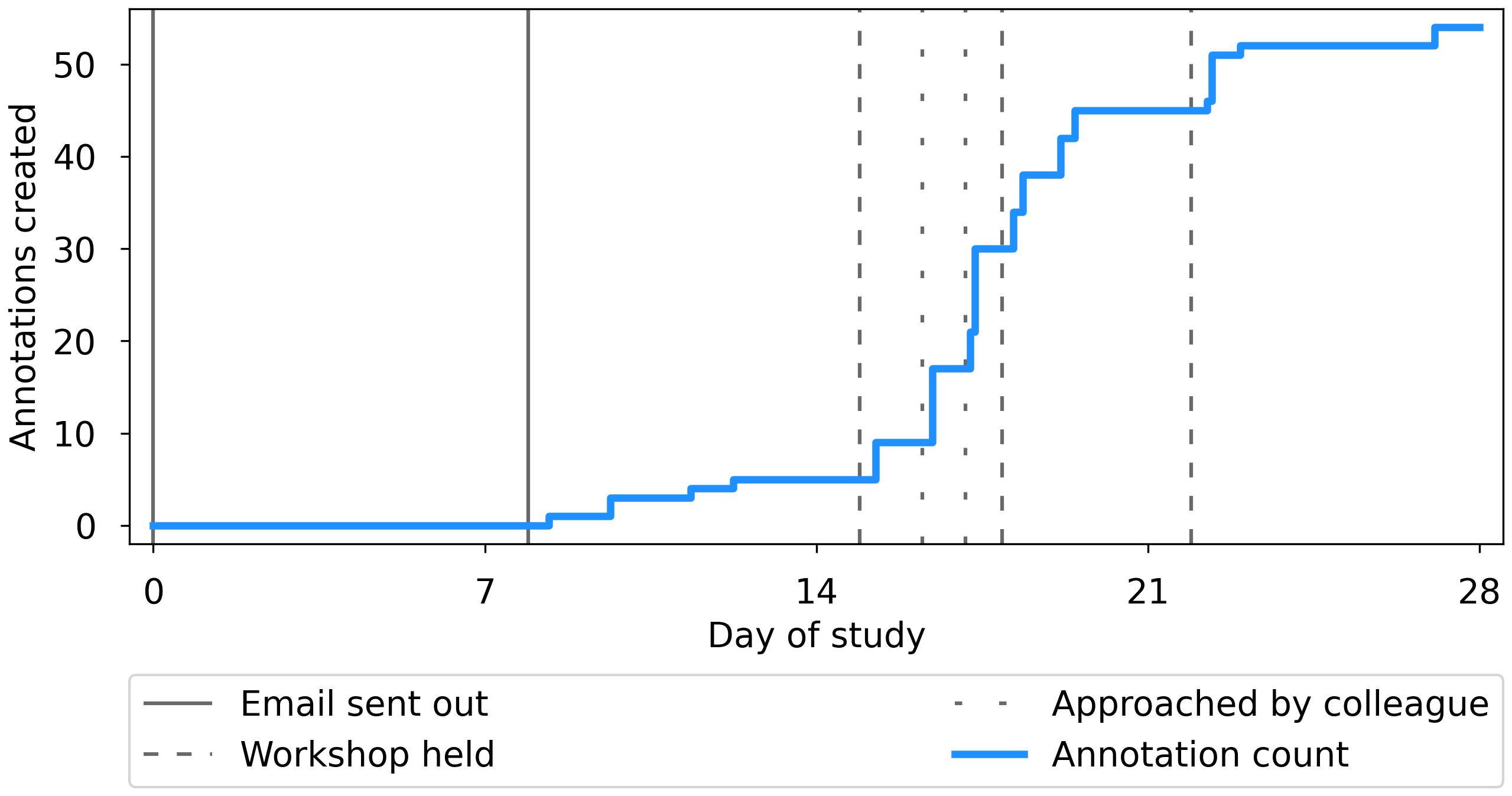}
        \caption{Cumulative annotation count throughout \ac{S1}. The initial lack of annotations has improved following the e-mail intervention and workshops, but the approach by the colleagues resulted in the largest count of annotations created using our tool.}
        \label{fig:s1_annotations}
    \end{subfigure}
    \caption{Participant engagement throughout \ac{S1} following the implemented interventions.}
    \label{fig:s1_interventions_annotations}
\end{figure}

We observed varying effects on the participants' involvement in the study following different types of interventions we carried out.
Overall, the most effective method of engaging with participants (generating consent form signatures, registrations and participation in the annotation task) was hosting the workshops in the \ac{ICU} \seeref{fig:s1_interventions}.
While the majority of the participants signed the consent forms and registered for an account following e-mail and workshop interventions, the primary motivators for engagement with the task were both workshops and the approach of the colleague.
In-person workshops which focused on both improving recruitment and promoting engagement with the annotation task proved to be effective but had short-lived effects in engaging participants with the task.
Addressing the prospective participants during workshops allowed for the recruitment of participants who did not respond to the previous e-mails but was limited by the availability of staff present when the workshop was taking place.
Utilising the approach of the colleague as a strategy appeared to be primarily effective in engaging existing participants rather than recruiting new ones, and resulted in the largest number of annotations being created, suggesting that utilising social connections to promote interest in the study can be particularly effective at improving engagement.

Approach by colleague led to the largest number of created annotations following the intervention \seeref{fig:s1_annotations}; despite being ineffective as a recruitment strategy, it had the highest conversion rate of participants who consented to the study, registered on the tool and annotated data.
Recruitment via e-mail resulted in participants registering on the system but not engaging with the annotation task \seeref{fig:s1_interventions}.

\subsection{Ethics Statement}
This work was approved by the Faculty of Engineering Research Ethics Committee at the University of Bristol (case 2022-150).

\section{Results}

\subsection{Feedback Questionnaire}

Participants found the interface of the data annotation tool to be intuitive and easy to use.
They expressed a preference for the graphical interface for creating rulesets and suggested that the node environment allowed them to define the logic in an intuitive way.
Participants also reflected on the onboarding experience and found it helpful as an introduction to the tool.
In their experience, they felt competent at using the tool after an initial 20-30 minutes of experimentation.
One of the participants suggested that an instructional video or an integrated tutorial could be helpful if the tool were to be scaled to a larger population.
Furthermore, participants felt that the process of iterating on the ruleset definition was intuitive and using the statistics page displayed following the ruleset creation helped them in refining the ruleset.
Some of the participants also found the ability to preview the annotations in the context of individual admissions to be valuable and helpful in identifying edge cases, but felt that it could be further improved to benefit the iterative process of creating rulesets.

The responses provided many suggestions for improvements to the tool, including the ability to compare the results of their annotations to those of other annotators, limiting the set of parameters to choose from when creating rules, facilitating a way to visually describe the impact of changes made between different rulesets and ability to specify the temporal change in categorical variables (e.g. change from a certain ventilator mode to a different one).
They also found themselves questioning the specificity of the annotation task and indicated that a concrete and robust definition of the research question and desired annotation is critical for the effectiveness and accuracy of the process.

Some of the challenges experienced during the task included having to account for a large number of proprietary modes of ventilation and not being able to access additional contextual information such as the reasons for admission, clinical notes and variables related to the medication intake.
Additionally, participants reported inconsistencies in the dataset which manifested as conflicting values for different parameters and a large variability in clinical practice present throughout the data, making it difficult to account for the edge cases.
They reflected on the time-consuming nature of the manual annotation process and the limitations associated with annotating large datasets using this approach.
Conversely, participants also found the semi-automated annotation to be time-efficient and described it as a feasible approach to time-series data annotation in the clinical setting.
One of the participants highlighted the importance of a close working relationship between the researchers and annotators and the role it can play in the effectiveness of the tool.

\subsection{Annotation}

\subsubsection{Individual annotation}

Over the course of 28 days, 12 participants annotated 31 unique admissions of which 4 were annotated by more than one participant, resulting in a total of 47 annotated admissions and an average of 3.92 admissions annotated per participant.
The admissions annotated by more than one participant were annotated by 9, 5, 4 and 2 distinct participants respectively.
118 individual annotations were created in total across all admissions and participants, resulting in an average of 2.51 annotations per admission and 9.83 annotations per participant.

\begin{figure}[tbp]
  \includegraphics[width=\textwidth]{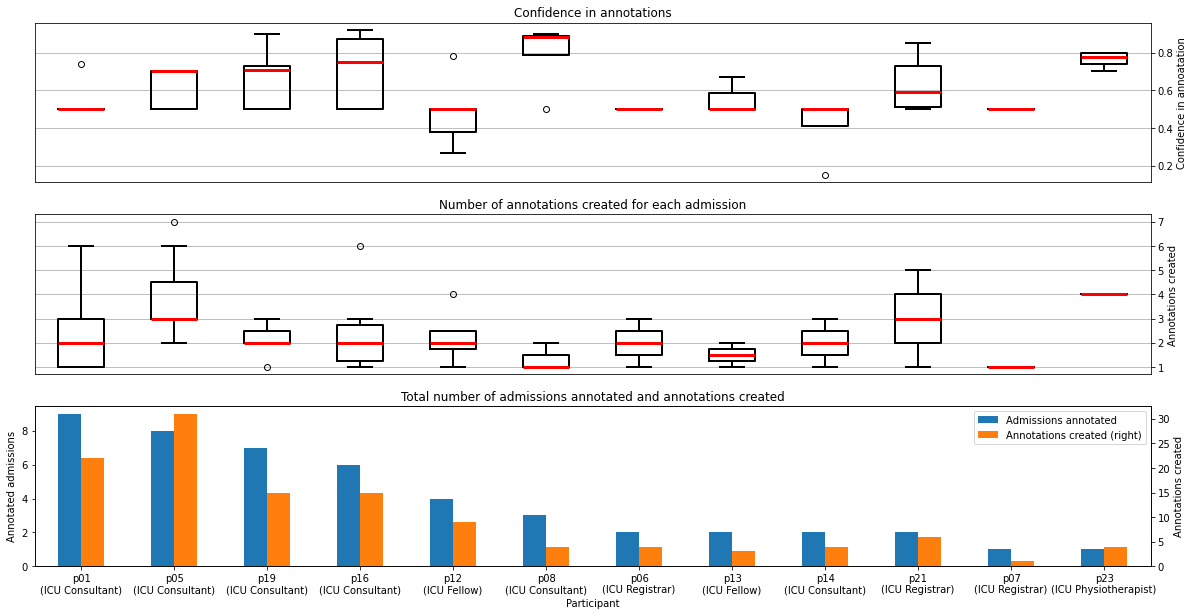}
  \caption{Characteristics of annotations created by different participants; the boxes encapsulate the \ac{IQR} as defined by Q3-Q1, the whiskers range up until Q1 - 1.5 \ac{IQR} and Q3 + 1.5 \ac{IQR}, outliers are plotted as individual circles and the median is plotted in red. The majority of participants stated their confidence as higher than 50\%, and the median number of annotations created for each admission was equal to or below 3 for the majority of the participants.}
  \label{fig:individual_annotation_characteristics}
\end{figure}

During the annotation process, participants could create any number of annotations for each admission and specify the confidence in the accuracy of each of the created annotations using a slider ranging between (0,1), with a default position of 0.5.
The majority of participants (8 out of 12) defined the confidence in the created annotations as either 0.5 or above, suggesting high overall confidence in the accuracy of the label.
The median number of annotations created for each of the admissions was equal to or lower than 3 for the majority of participants (11 out of 12).
The characteristics of annotations created by the participants during \ac{S1} are depicted on the \autoref{fig:individual_annotation_characteristics}.

\begin{figure}
    \centering
    \begin{subfigure}[t]{0.49\textwidth}
    \includegraphics[width=\textwidth]{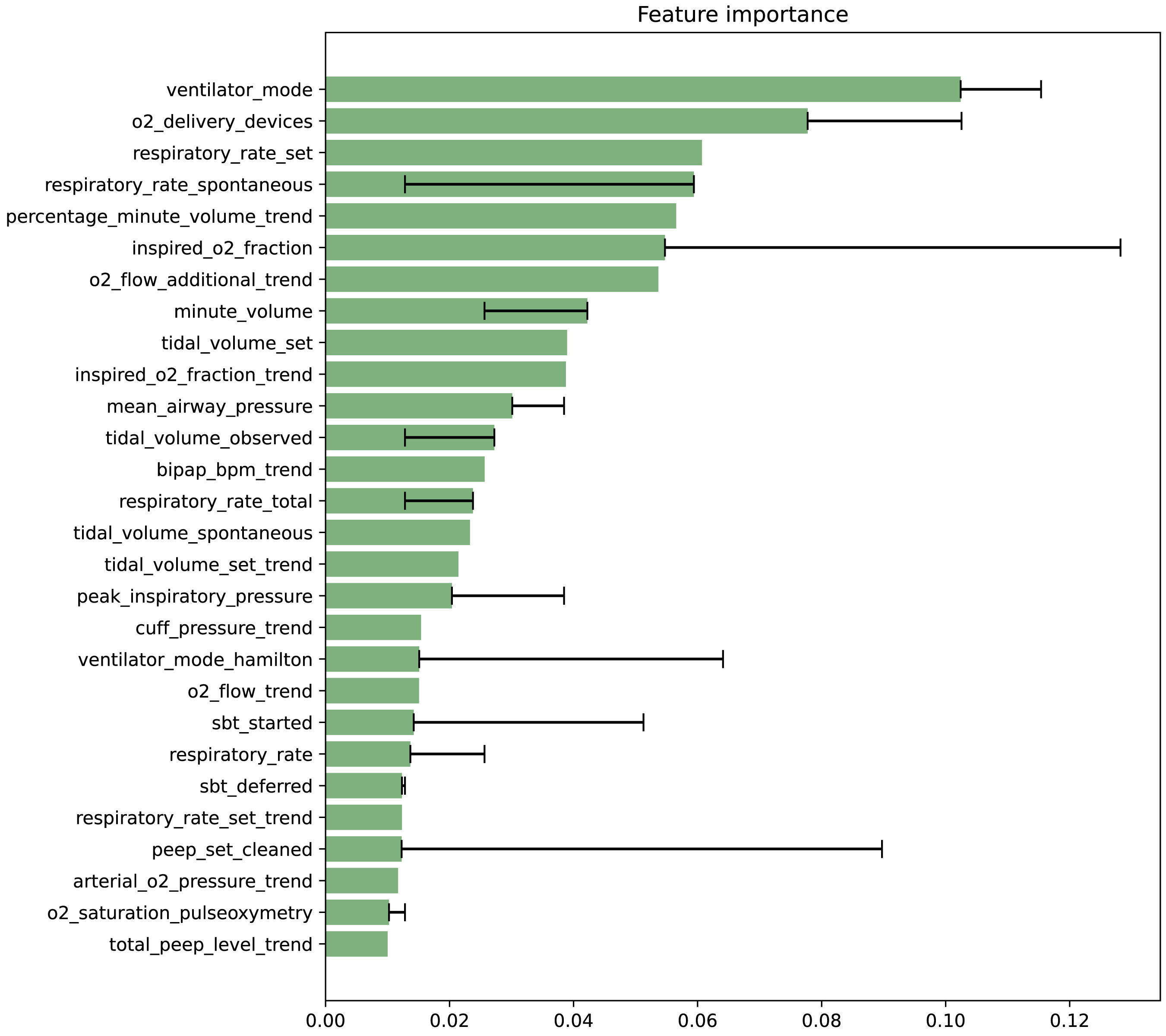}
    \caption{}
    \label{fig:individual_annotation_decision_tree_a}
    \end{subfigure}
    \begin{subfigure}[t]{0.49\textwidth}
    \includegraphics[width=\textwidth]{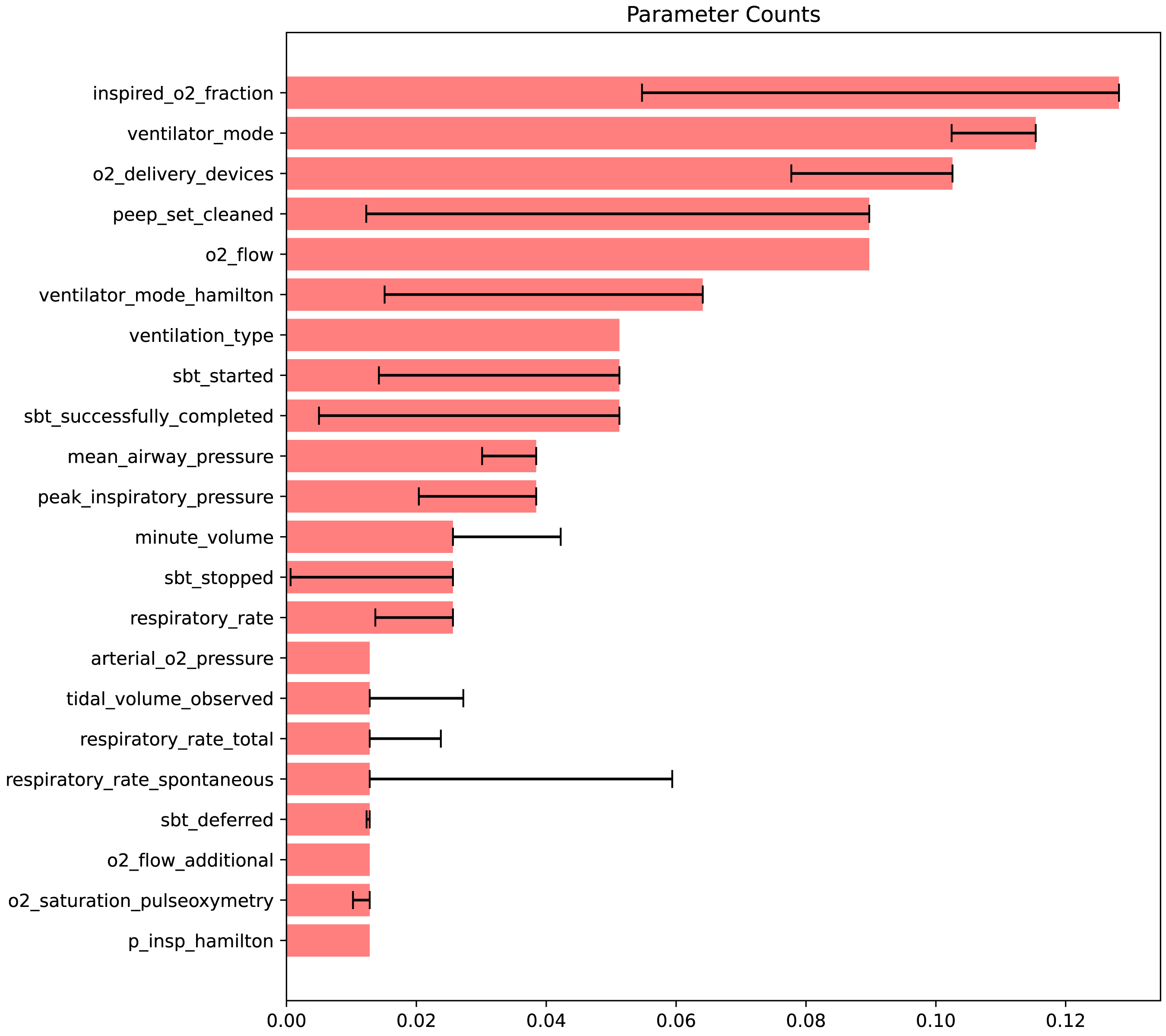}
    \caption{}
    \label{fig:individual_annotation_decision_tree_b}
    \end{subfigure}
\caption{\autoref{fig:individual_annotation_decision_tree_a} depicts the feature importance for the decision tree classifier trained from \ac{S1} annotations; features with importance $<1\%$ are discarded. \autoref{fig:individual_annotation_decision_tree_b} depicts the parameter prevalence as described by the count of participants who used that parameter in the individual annotations over a sum of all unique participant-parameter pairs used. The error bars describe the distance between the importance of the feature in the decision tree and the prevalence of a given parameter in individual annotations.}
\label{fig:individual_annotation_decision_tree}
\end{figure}

To gain a better insight into the collective approach to annotation of multiple annotators, we trained a decision tree classifier based on the labels created by all \ac{S1} participants.
The model assumed two classes (\textit{weaning} and \textit{not weaning}) which were computed using the majority threshold for each of the annotated admissions.
To establish optimal parameters for the classifier we ran a 10-fold grid search hyper-parameter tuning across a maximum number of features and maximum tree depth ranging between (1,100) and (1,100) respectively.
We reached an accuracy of 0.975 under a 70\% train, 30\% test split with the optimal parameters of 36 maximum features and a maximum depth of 52.
A comparison of the parameters specified by the participants as part of the annotations created during \ac{S1} and the features of the decision tree classifier yielded several parameters that placed high on both lists, including ventilator mode, oxygen delivery devices and the fraction of inspired oxygen.
The feature importance of the decision tree and the prevalence of parameters used by the participants in \ac{S1} are depicted on the \autoref{fig:individual_annotation_decision_tree}.

\subsubsection{Semi-automated annotation}

\begin{figure}[tbp]
  \includegraphics[width=\textwidth]{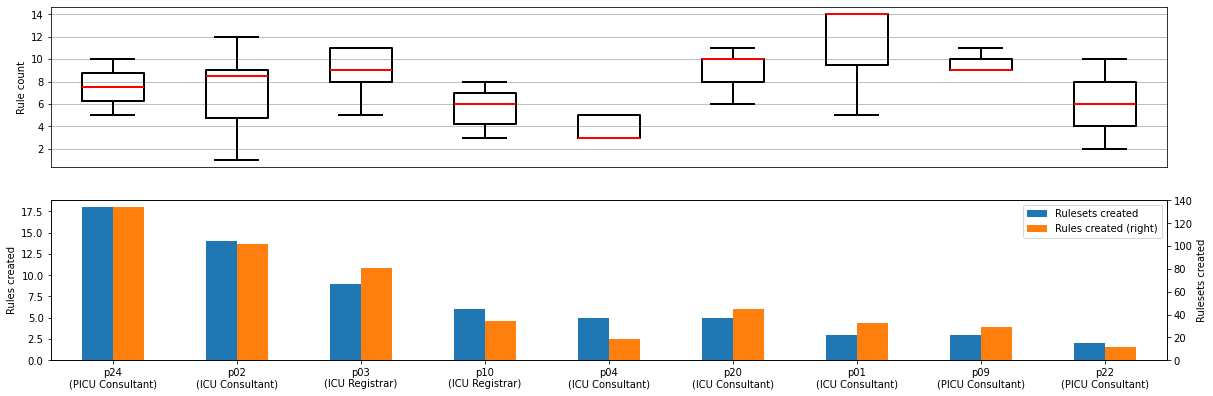}
  \caption{The median count of rules within a ruleset was between 5 and 10 for the majority of participants. The boxes encapsulate the \ac{IQR} as defined by Q3-Q1; the whiskers range up until Q1 - 1.5 \ac{IQR} and Q3 + 1.5 \ac{IQR}; the median is plotted in red.}
  \label{fig:ruleset_boxplot}
\end{figure}

The goal of \ac{S2} was for each participant to design a ruleset that, to the best of their ability, annotates weaning from mechanical ventilation when applied to the data.
During \ac{S2}, 9 participants created a total of 65 unique rulesets composed of 489 rules and 255 relations.
This resulted in an average of 7.22 rulesets, 54.33 rules and 28.33 relations per participant and 7.52 rules and 3.92 relations per ruleset.
Each ruleset was evaluated against all 1,967,145 individual hours of admissions in the dataset with an average execution time of 1 minute and 20 seconds.
This produced a total number of 2,421,873 annotations across all rulesets and 763,581 annotations across the final rulesets for each participant.
The distribution of rule counts across rulesets for each of the participants is depicted in \autoref{fig:ruleset_boxplot}.

\begin{figure}[tbp]
  \includegraphics[width=\textwidth]{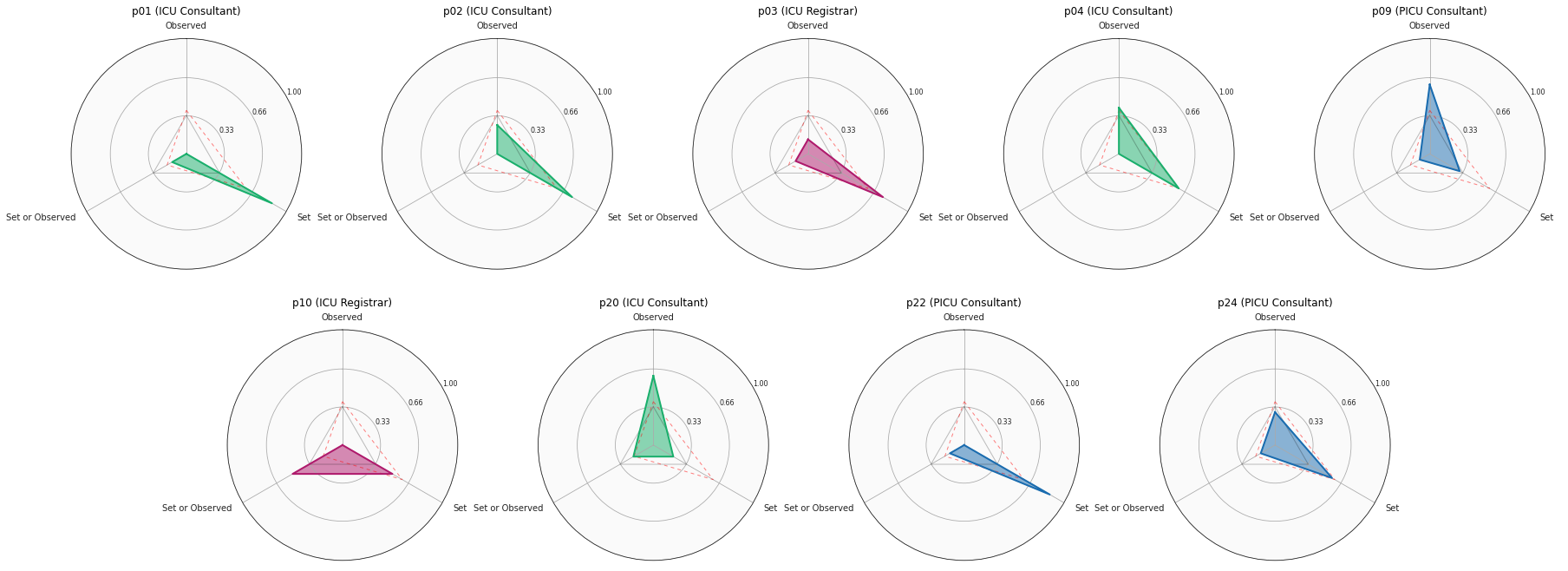}
  \caption{Radial charts depict the preference for different types of parameters across the participants' final rulesets; the average is depicted in a dashed line. We observed different preferences among participants where preference was defined as a parameter type with the highest prevalence. 7 of the participants preferred \textit{set parameters} over other types, and 2 participants preferred \textit{observed parameters}.}
  \label{fig:ruleset_types_of_parameters}
\end{figure}

We observed a large variance across the parameters used in the rulesets to annotate the data.
The parameters were aggregated based on their type into three categories including \textit{set parameters} explicitly defined by the clinician (e.g. mode of ventilation), \textit{observed parameters} measured for the patient (e.g. heart rate) and \textit{set-or-observed parameters} which could change depending on the characteristics of treatment (e.g. pressure delivered by the ventilator, which is set in pressure-controlled modes of ventilation, but observed in the volume-controlled modes of ventilation).
Participants exhibited different preferences in parameter types they used in their final rulesets \seeref{fig:ruleset_types_of_parameters}.
On average 64\% of rules used in the final rulesets were \textit{set parameters}, 24\% were \textit{observed parameters} and the remaining 12\% were \textit{set-or-observed parameters}.
3 of the participants used only \textit{set} and \textit{set-or-observed} parameters and 3 other participants used only \textit{set} and \textit{observed} parameters; the 3 remaining participants used all three types of parameters.
We found no correlation between the roles of participants and their preference towards different types of parameters.

\subsubsection{Approach comparison}

\begin{figure}[tbp]
  \includegraphics[width=\textwidth]{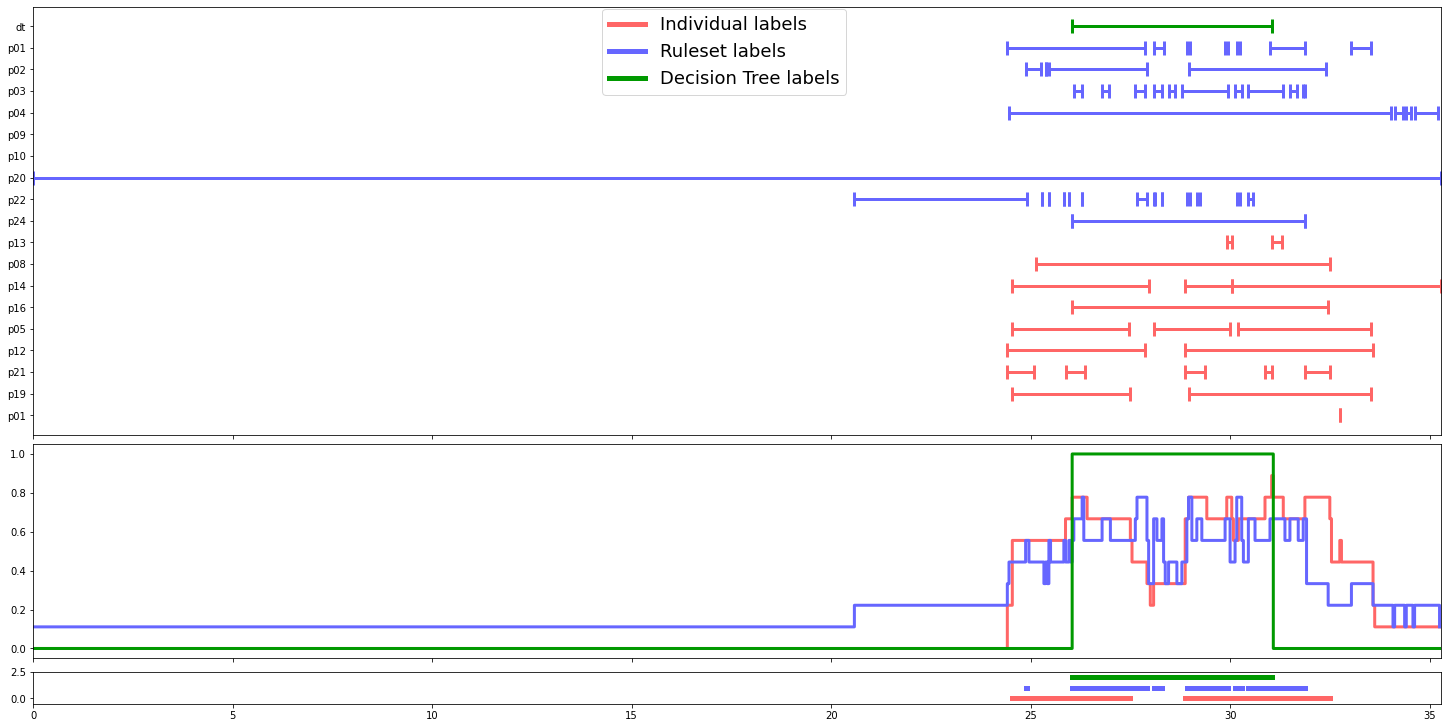}
  \caption{Annotations created using different techniques presented a large overlap on the admission annotated by the highest number of annotators during \ac{S1}. \autoref{fig:most_annotated_admission}a depicts the timeline for the admission, together with the annotations created by different participants, aggregated by annotation type. \autoref{fig:most_annotated_admission}b compares the mean number of annotations present across the admission timeline, aggregated by annotation type. \autoref{fig:most_annotated_admission}c compares the different annotation types using the majority of annotators as a threshold for annotation presence (mean $> 0.5$).}
  \label{fig:most_annotated_admission}
\end{figure}

We compared the annotations in the context of the single admission annotated by the highest number of annotators during \ac{S1} (9 annotators) to establish the similarity between different approaches to annotation.
This comparison included annotations created individually during \ac{S1}, the annotations created by participants using rulesets during \ac{S2} as well as those created using a ruleset derived from the decision tree classifier trained on the individual annotations from \ac{S1} \seeref{fig:most_annotated_admission}.
We observed the smallest average error between annotations created during \ac{S1} (12.28\%) and the highest average error between annotations created during \ac{S2} (30.65\%) \seeref{table:average_error_single_admission}.
The average error among annotations created during both stages was 28.47\% and aggregating the annotations by type and applying a majority threshold of $>$ 50\% produced an average error of 5.19\% for annotations created by both stages \seeref{table:average_error_single_admission}.
For this particular admission, we also observed differences between the continuity of annotations across different approaches.
The decision tree-based ruleset provided a single annotation, \ac{S1} annotations provided an average of 2.11 annotations per participant and \ac{S2} annotations provided an average of 4.33 annotations per participant.

\begin{table}
\setstretch{1.25}
\caption{Average error between annotations of different types for admission annotated by the highest number of annotators during \ac{S1}. The average error was computed as the average number of mismatched annotations between all admission hours across all participants.}
\begin{tabularx}{\textwidth} {
  | >{\hsize=.4\textwidth\raggedright\arraybackslash}X 
  | >{\raggedleft\arraybackslash}X
  | >{\raggedleft\arraybackslash}X | }
 \hline
 \multirow{2}{*}{\textbf{Annotation source}} & \multicolumn{2}{c|}{\textbf{Average error [\%]}} \\
 \cline{2-3}
 & \multicolumn{1}{c}{\textbf{Unaggregated}} & \multicolumn{1}{|c|}{\textbf{Threshold $>$ 50 \%}} \\
 \hline
 Stage 1 & 12.28 \% & - \\
 \hline
 Stage 2 & 30.65 \% & - \\
 \hline
 Stage 1 and Stage 2 & 21.90 \% & 5.19 \% \\
 \hline
 Stage 1 and Decision Tree & 12.29 \% & 12.28 \% \\
 \hline
 Stage 2 and Decision Tree & 28.47 \% & 7.56 \% \\
 \hline
 Stage 1, Stage 2 and Decision Tree & 21.32 \% & 4.17 \% \\
 \hline
\end{tabularx}
\label{table:average_error_single_admission}
\end{table}

\begin{figure}[tbp]
  \includegraphics[width=\textwidth]{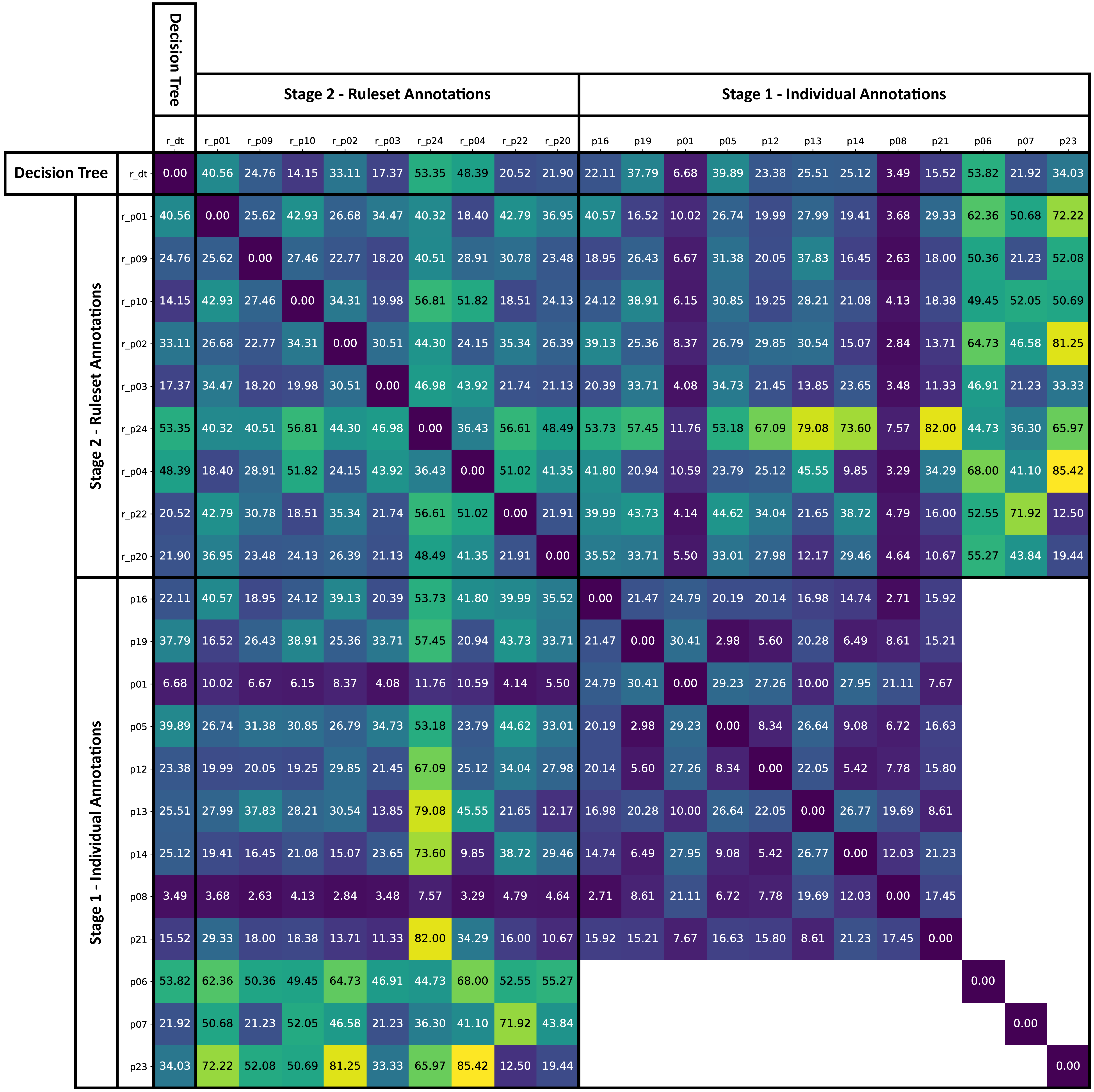}
  \caption{Confusion matrix for annotations created individually during \ac{S1} and semi-automated annotations created using rulesets during \ac{S2} as well as those created by the ruleset built to resemble the decision tree trained on the individual annotations from \ac{S1}. The difference between annotations from different sources is expressed in percentages, where 0\% denotes perfect match and 100\% denotes lack of overlap. Missing data indicates that specific annotation sources did not share any admissions.}
  \label{fig:confusion_matrix}
\end{figure}

Aggregating the annotations created using each approach across all shared admissions allowed us to establish the similarity between different participants and across different approaches \seeref{fig:confusion_matrix}.
The mean error across annotations created during \ac{S1} was 15.94\%, those created during \ac{S2} - 33.78\% and across both \ac{S1} and \ac{S2} annotations - 31.62\% \seeref{table:average_error_entire_dataset}.
We observed a particularly low error (2.98\%) between \ac{S1} annotations created by participants p19 and p05 and a minimum error (2.63\%) between \ac{S1} annotations of p08 and \ac{S2} annotations of p02.
The annotations created by the participant who took part in both stages (p01) achieved an error of 10.02\% between \ac{S1} and \ac{S2} annotations.
In particular, annotations created by the participants p08 and p01 in \ac{S1} exhibited the lowest average error when compared to \ac{S2} annotations and the decision-tree-based annotations.

\begin{table}[ht]
\setstretch{1.25}
\caption{Average error between annotations of different types for across the entire dataset. The average error was computed as the average number of mismatched annotations between all admission hours across all participants.}
\begin{tabularx}{\textwidth} { 
  | >{\hsize=.4\textwidth\raggedright\arraybackslash}X 
  | >{\raggedleft\arraybackslash}X | }
 \hline
 \textbf{Annotation source} & \multicolumn{1}{|c|}{\textbf{Average error [\%]}} \\
 \hline
 Stage 1 & 15.94 \%\\
 \hline
 Stage 2 & 33.78 \%\\
 \hline
 Stage 1 and Stage 2 & 31.62 \% \\
 \hline
 Stage 1 and Decision Tree & 25.77 \% \\
 \hline
 Stage 2 and Decision Tree & 30.46 \% \\
 \hline
 Stage 1, Stage 2 and Decision Tree & 28.80 \% \\
 \hline
\end{tabularx}
\label{table:average_error_entire_dataset}
\end{table}

\section{Discussion}

\subsection{Effectiveness of Interventions}

Recruitment of research participants can be a challenging process that evolves over the course of the research to adapt to the changing circumstances and conditions \cite{bonisteel_reconceptualizing_2021, negrin_successful_2022}.
Despite the efforts made to accommodate the busy schedules of the clinical staff working in the \acp{ICU}, such as asynchronous and remote enrolment and annotation, we encountered challenges in recruiting and effectively engaging participants during \ac{S1} of our study.
The recruitment e-mails we sent out failed to generate a response that satisfied our criteria, which could be attributed to several factors.
The task brief asked participants to annotate as many responses as they could in the time frame provided, suggesting a higher time commitment than that of \ac{S1} and therefore lack of engagement with the study.
Furthermore, the asynchronous and remote nature of the task meant that participants were unable to seek additional support, particularly at the early stage of the activity.
Finally, the lack of engagement with the study beyond granting consent or registration could be attributed to the inconsequential nature of this remote participation and the lack of incentive that would motivate further commitment to the task.

We observed improved conversion of enrolled to active participants, as well as rise in engagement with the annotation activity following the implementation of several interventions.
Resending the invitational e-mail prompted the existing participants to create an account in the tool and undertake the annotation task, but failed at recruitment of new participants.
This suggests a follow-up communication to be a viable strategy for re-engaging existing participants who have already enrolled in a study, but also highlights the limitations of repeated attempts at recruitment via e-mail.
A technique that proved effective in recruiting participants involved directly incentivising prospective participants by hosting workshops in the form of structured learning events.
These events were required by the clinical staff in the trainee roles to progress their careers and their deployment resulted in the most effective acquisition of new participants in our study.
Conversely, this technique had a limited effect on generating prolonged engagement following the workshops, which could be attributed to the high frequency at which research is being conducted at this particular hospital and therefore research fatigue.
We also anticipate that this particular incentive was of limited appeal to participants who were not trainees and that offering a more ubiquitous incentives could prove more effective at recruiting more diverse population.
Finally, incorporating one of the members of clinical staff from the research site within our research team proved to be the most impactful technique for promoting engagement with the study activity.
We anticipate that this form of utilising the existing community at the hospital helped build trust and credibility of our research, and allowed us to approach the participants on a more personal level, which in turn encouraged their involvement in the study.
While we observed substantial improvements in participation following this type of intervention, we acknowledge that it may not be suitable in certain settings, particularly when there is a lack of existing connection between the facilitators and the research site.

\subsection{Usability}

Data annotation is a complex and time-consuming process that relies on the domain expertise of annotators for bringing in additional context to the data.
Within intensive care settings, where the number of available annotators is very limited, their time is in high demand and the volumes of data are frequently too large for being labelled directly, annotating data poses a substantial challenge.
This calls for a solution that is both effective and time-efficient, offering a workflow that minimises the friction with the tool and maximises the productivity of the annotator.
The qualitative findings on the usability of our semi-automated annotation feature suggested that participants were able to use the software effectively and found the interface intuitive and easy to use.
In their feedback, participants indicated a preference towards the semi-automated approach over the direct annotation of individual admissions and reflected on the time-consuming nature of the manual annotation of data, speaking to the appeal of using the graphical interface for creating rulesets.
These results are supported by our quantitative analysis where participants were able to define the rulesets that describe how annotations should be applied and observe the results of the annotation in an average of 1 minute and 20 seconds (in addition to the time spent on describing the ruleset using our graphical interface).
Furthermore, when questioned about the practicality of semi-automated annotation for large clinical datasets, participants identified this approach as a more feasible technique under the limiting circumstances of intensive care settings.
User acceptance plays a critical role in the successful adoption of software \cite{taherdoost_review_2018,momani_technology_2017}, which is particularly important in the context of clinical users annotating medical data.
Ensuring that the provided functionality meets the standards of the end users can help prevent errors when using the software and ultimately lead to better outcomes \cite{damodaran_user_1996, slattery_research_2020, teixeira_using_2007, martin_user-centred_2012, davey_requirements_2015, ocathain_guidance_2019}.

\subsection{Requirements for Semi-automated Annotation}

In our previous research, we established several requirements for the digital data annotation tool that could be used in the intensive care setting \cite{wac_capturing_2023}.
In that study we observed multiple annotators labelling individual admissions and analysed their actions to gain insight into the process of interaction with the annotation task.
As part of that approach, we observed participants reflecting on the data through the lens of the previously created annotations, allowing them to better understand the data and evaluate their annotations, frequently resulting in adjustments that improved the quality of their annotations \cite{wac_capturing_2023}.
We identified this as a key characteristic of the annotation process and hypothesised that, while this observation was made based on the annotation of individual admissions, the reflexive process itself would also apply in a wider annotation context, including the semi-automated approach to annotation.
Consequently, we reflected it in our requirements as a need for the ability to investigate and analyse the created annotations.
In our current study, participants described the ability to analyse the annotations created using rulesets as valuable and helpful in refining their rulesets and improving the accuracy of their annotations.
These findings support our initial hypothesis and highlight the reflexive nature of annotation and the importance of its capture in the digital tool, irrespective of the mechanisms used to annotate the data.

\subsection{Approach to Annotation}

The analysis of the data collected during both stages of this study allowed us to quantitatively assess the feasibility and limitations of the semi-automated annotation in the clinical setting.
Annotations created by participants during \ac{S1} were characterised by high confidence in the annotations overall, however, we observed this metric to be particularly high for \ac{ICU} Consultants in comparison to other roles.
This could be attributed to the seniority of the position (in comparison to the \ac{ICU} Registrars), as well as the responsibilities associated with this job role.
Collecting the self-described confidence metric supplementing the annotation is an important aspect of crowdsourced data annotation that can support the later use of annotations in \ac{ML} contexts \cite{ni_understanding_2013}.

To enhance the understanding of the annotation process, we trained a decision tree classifier based on the annotations created during \ac{S1}.
The analysis of the feature importance of that model and the parameters marked as relevant during the individual annotation in \ac{S1} allowed us to cross-validate the importance of different parameters on the specific annotation task of our study -- weaning from mechanical ventilation.
In our analysis, several parameters scored high across both lists, including ventilator mode, oxygen delivery devices and the fraction of inspired oxygen, suggesting the key role of these parameters in the weaning process and our desired label.
We observed an average error of 15.94\% within annotations created during \ac{S1}, suggesting some degree of disagreement between the annotators and likely an inherent uncertainty within the task.
Together with the high accuracy of our model and the relatively small difference to a low average error between \ac{S1} annotations and those created using our model (25.77\%), our findings suggest the feasibility of a decision tree model for several use-cases in the annotation process.
Firstly, the model could be used as a preliminary measure to establish the conformity of individual annotators to the overall annotators group by capturing the distance between their individual annotations and those created by the decision tree classifier.
Calculating the divergence of annotators in specific cases could be used for anomaly detection and in consequence \say{active learning}, in which annotations of particular disagreement would be highlighted and prioritised for additional input by annotators.
Secondly, the explainable nature of the decision tree classifier could also be used to aid the interpretability of the resulting model by providing a set of traceable steps used to create the labels.
In the context of machine learning, and in particular the critical settings such as intensive care, explainability plays a critical role in  decision support systems as it allows the clinical staff to understand the reasoning behind suggested decisions, which in turn builds trust and provides safeguarding mechanisms necessary within healthcare.

By analysing the final rulesets created by each of the participants and aggregating the parameters they used in their rulesets, we were able to identify different preferences within our participant population.
The majority of the participants (7 out of 9) chose to use primarily the \textit{set parameters}, while the 2 remaining participants focused primarily on the \textit{observed parameters}.
Despite the fact that we did not observe the correlation between the choice of different parameter types and participants' job roles, this analysis generated substantial insights into the process of annotation within the clinical setting.
Primarily, the use of different types of parameters could be attributed to several factors including the desired subject of the annotation, the preferences of annotators in how they assess the patients and the combination of both.
Treatment or delivery of specific procedures and interventions is informed by the condition the patient is in and their individual needs.
While these needs could be assessed by focusing on the \textit{observed parameters}, gaining a broader understanding of what led to this condition in the first place may also require analysis of the \textit{set parameters} (e.g. high heart rate (\textit{observed parameter} could be attributed to a specific mode of ventilation (\textit{set parameter}) that a patient is on).
Furthermore, creating labels that focus on a treatment-oriented task (e.g. presence of weaning from mechanical ventilation) will likely result in focus on the type of parameters which describe the delivery of that treatment -- the \textit{set parameters}.
Conversely, labels capturing the inherent patient condition (e.g. dependence on the ventilator) could instead focus on the \textit{observed parameters} which describe the overall state of the patient and their response to treatment.
This insight is significant for the process of data annotation because it allows for prioritisation of the data cleaning efforts and selection of relevant parameters depending on the annotation task, but it also informs the inherent biases that may be present in the created labels.
\textit{Set parameters} specific to a given the treatment are a result of the clinical decisions made at the time when patient care was delivered.
This suggests that both the way in which the care was delivered and the resulting patient condition can be biased towards specific treatment approaches or characteristics of the data collection site, which should be taken into account when collecting the data annotations.
Lastly, the choice to use different parameter types can provide insight into the thought process and approach to data annotation of specific annotations which could further enhance the collected annotations.
Annotators who focus on any specific subset of parameters may inherently omit other data sources and fail label cases which don't align with their thought process.
Alternatively, their bias towards certain parameters could also translate to the created labels and influence the performance of the model trained on those annotations.

\subsection{Strengths and Limitations}

Our study focused on evaluating the digital time-series data annotation tool purpose-built for the clinical settings.
We explored the feasibility of a novel, semi-automated approach to creating annotations, which facilitated time-efficient annotation of large volumes of data by a limited number of annotators characteristic of the intensive care settings.
In this study, we demonstrated the usability and feasibility of collecting annotations using our tool in a real-world scenario using real patient data.
To the best of our knowledge, this is the first study of its kind focused on evaluating the semi-automated approach to collecting annotation within the clinical setting.
One of the strengths of this study was the use of mixed methods for the evaluation, which allowed us to capture both objective quantitative data on the use of the software and more detailed and personal insights from the participants.
Utilising mixed methods and different sources of data is a strategy that provides a holistic picture of the results and improves the trustworthiness of research.
In our findings, the collected results supported our previous hypothesis on the importance of re-evaluating data in the context of created annotations, which was reflected in both the number of created rulesets and annotations and the data collected via feedback questionnaires.

While the low overlap of participants between \ac{S1} and \ac{S2} did not prevent us from evaluating the tool, it limited our ability to measure the differences between annotations created directly on data and those created via rulesets within the context of individual annotators.
We anticipate that this type of analysis could benefit our study and further improve the understanding of the limitations of our tool and the characteristics of the annotation process as a whole.
Furthermore, this study focused on collecting data annotations explicitly in the context of weaning from mechanical ventilation, which resembled the task definition from our previous study.
This allowed us to directly evaluate the relevance of the captured requirements without introducing the bias of a different task, but limited the generalisability of our results as it likely introduced a bias towards that specific task.

\subsection{Future Work}

The primary goal of the our evaluation was aimed at identifying the strengths and limitations of the proposed design, assessing its usability and identifying potential improvements to the tool.
To that extent, we conducted a data annotation activity and collected annotators' input using two distinct techniques -- by directly annotating individual admissions and by creating rulesets which could be used to annotate the data.
Future work should therefore focus on evaluating the performance of the annotations collected via our tool within a broader machine learning contexts.
This could include addressing the problem of modelling the accuracy of annotations created by multiple experts, particularly when the ground truth is not unavailable \cite{ratner_data_2016}, as well as exploring the techniques for aggregating annotations to establish more robust training datasets.
In particular, while the rulesets collected using our tool were used to directly annotate the data in order to allow for their analysis and refinement, research should focus on aggregating multiple rulesets to develop a more durable annotation model.
To that extent, we suggest that the development of such model follows the approaches previously demonstrated in weak supervision techniques such as data programming \cite{ratner_data_2016, snorkel_team_labelmodel_2020}, which are particularly suitable given the similarities between the rulesets and labelling functions.
Furthermore, we anticipate that incorporating additional functionality that facilitates active learning within our tool would result in further improvements to the efficiency of the annotation process and provide unique opportunities for effective annotation within the clinical setting.

Finally, the evaluation of our tool at a larger scale and among a wider population of annotators could further improve the understanding of the feasibility and limitations of our proposed tool.
In particular, the effectiveness of the tool could be measured in a study that defines an annotation task surrounding a concrete problem with a pre-established ground truth and with a higher overlap of participants between different annotation methods.
Such study could provide further insights into the inter-annotator differences, their needs and preferences, and strengthen the understanding of the annotation process within the clinical setting.
While our tool was designed to support asynchronous annotation by a large number of geographically distributed users, the evaluation took place in a single research site over a limited span of time.
Conducting a multi-centre study would therefore benefit the evaluation of our tool and could also serve as a basis for identifying the differences and variation in practice between different sites.

\subsection{Conclusions}

In this study, we presented a digital time-series data annotation tool for clinical settings.
We demonstrated its use in practice using real patient data from an anonymised dataset with expert annotators from \acp{ICU} and evaluated its usability.
By conducting two stages focused on individual and semi-automated annotation, we explored how clinicians approach the task of data annotation and investigated the feasibility of the semi-automated approach facilitated by our tool.
The results of our study demonstrated satisfaction with the tool and suggested the feasibility of its use for the annotation of large clinical datasets.
We observed similarities in annotations created using different techniques, discussed the challenges of opportunities of the annotation in the healthcare setting and provided suggestions for future research directions.

\clearpage

\subsection*{Acknowledgements}

This work was partly supported by the Engineering and Physical Sciences Research Council Digital Health and Care Centre for Doctoral Training at the University of Bristol (UKRI grant EP/S023704/1 to MW).
CMW was funded by the South West Better Care Partnership, supported by Health Data Research UK.
The project received an \ac{AWS} Cloud Credit for Research grant.

\subsection*{Author Contributions}

MW was involved in the study design, development of the software, data collection, data analysis, and writing.
CMW and RSR were involved in study design and critical revision.

\subsection*{Conflicts of Interest}
None declared.

\subsection*{Acronyms}

\begin{acronym}
 \setlength{\parskip}{0ex}
 \setlength{\itemsep}{-1ex}
 \setstretch{1}
 \acro{AWS}{Amazon Web Services}
 \acro{ICU}{Intensive Care Unit}
 \acro{IQR}{Interquartile Range}
 \acro{MIMIC}{Medical Information Mart for Intensive Care IV version 2.0}
 \acro{ML}{Machine Learning}
 \acro{NHS}{National Health Service}
 \acro{S1}{Stage 1}
 \acro{S2}{Stage 2}
 \acro{SQL}{Structured Query Language}
 \acro{UHBW}{University Hospitals Bristol and Weston NHS Foundation Trust}
 \acro{UK}{United Kingdom}
\end{acronym}

\appendix

\clearpage

\section*{Appendix}

{
\renewcommand\tabularxcolumn[1]{m{#1}}
\newcolumntype{P}{>{\hsize=.2\hsize\centering\arraybackslash}X}
\newcolumntype{Q}{>{\hsize=.65\hsize}X}
\newcolumntype{S}{>{\hsize=.15\hsize\centering\arraybackslash}X}
\footnotesize
\setstretch{1}
\setlength\aboverulesep{1pt} \setlength\belowrulesep{1pt}
\setlength\extrarowheight{1pt}
\begin{xltabular}{\textwidth}{
|P|Q|S|
}
\caption{Parameters used for the annotation activity. \label{appendix:parameters}}\\
\hline
\textbf{Parameter Group} & \textbf{Parameter} & \textbf{Unit of measurement}\\
\hline
\multirow{12}*{Vitals} & Heart Rate & bpm \\ \cline{2-3}
& Arterial Blood Pressure Systolic & mmHg \\ \cline{2-3}
& Arterial Blood Pressure Diastolic & mmHg \\ \cline{2-3}
& Arterial Blood Pressure Mean & mmHg \\ \cline{2-3}
& Non-invasive Blood Pressure Systolic & mmHg \\ \cline{2-3}
& Non-invasive Blood Pressure Diastolic & mmHg \\ \cline{2-3}
& Non-invasive Blood Pressure Mean & mmHg \\ \cline{2-3}
& Cardiac Output (thermodilution) & L/min \\ \cline{2-3}
& Respiratory Rate & insp/min \\ \cline{2-3}
& Arterial $O_{2}$ Pressure & mmHg \\ \cline{2-3}
& Arterial $O_{2}$ Saturation & \% \\ \cline{2-3}
& $O_{2}$ Saturation (pulse-oximetry) & \% \\ \hline
\multirow{34}*{Respiratory} & PEEP Set & cm $H_{2}O$ \\ \cline{2-3}
& Temperature & C \\ \cline{2-3}
& PCWP & mmHg \\ \cline{2-3}
& $F_{i}O_{2}$ & \% \\ \cline{2-3}
& Ventilator Mode & \\ \cline{2-3}
& Cuff Pressure & cm $H_{2}O$ \\ \cline{2-3}
& Tidal Volume (set) & mL \\ \cline{2-3}
& Tidal Volume (observed) & mL \\ \cline{2-3}
& Tidal Volume (spontaneous) & mL \\ \cline{2-3}
& Minute Volume & L/min \\ \cline{2-3}
& Minute Volume Target (for IBW) & \% \\ \cline{2-3}
& Respiratory Rate (set) & insp/min \\ \cline{2-3}
& Respiratory Rate (spontaneous) & insp/min \\ \cline{2-3}
& Respiratory Rate (total) & insp/min \\ \cline{2-3}
& Peak Inspiratory Pressure & cm $H_{2}O$ \\ \cline{2-3}
& Plateau Pressure & cm $H_{2}O$ \\ \cline{2-3}
& Mean Airway Pressure & cm $H_{2}O$ \\ \cline{2-3}
& Total PEEP Level & cm $H_{2}O$ \\ \cline{2-3}
& SBT Started & \\ \cline{2-3}
& SBT Stopped & \\ \cline{2-3}
& SBT Succeeded & \\ \cline{2-3}
& SBT Deferred & \\ \cline{2-3}
& Expiratory Ratio & \\ \cline{2-3}
& Inspiratory Ratio & \\ \cline{2-3}
& Inspiratory Pressure (Draeger) & cm $H_{2}O$ \\ \cline{2-3}
& BiPAP Mode & \\ \cline{2-3}
& BiPAP EPAP & cm $H_{2}O$ \\ \cline{2-3}
& BiPAP IPAP & cm $H_{2}O$ \\ \cline{2-3}
& BIPAP BPM (S/T - back up) & bmp \\ \cline{2-3}
& $E_{t}CO_{2}$ & mmHg \\ \cline{2-3}
& Ventilator Mode (Hamilton) & \\ \cline{2-3}
& Inspiratory Pressure (Hamilton) cm $H_{2}O$ & \\ \cline{2-3}
& Resistance (expiratory) & cm $H_{2}O$/L/sec \\ \cline{2-3}
& Resistance (inspiratory) & cm $H_{2}O$/L/sec \\ \hline
\multirow{4}*{Equipment} & Ventilation Type & \\ \cline{2-3}
& Oxygen Delivery Devices & \\ \cline{2-3}
& Oxygen Flow & L/min \\ \cline{2-3}
& Oxygen Flow (additional) & L/min \\
\hline

\end{xltabular}
}

\bibliographystyle{ieeetr}
\bibliography{references}

\end{document}